\documentclass[12pt,preprint]{aastex}
\def\reference{\par\noindent\hangindent=1cm\hangafter=1}

\newcommand{\eq}{\begin{equation}}

\newcommand{\ff}[2]{{\textstyle \frac{#1}{#2}}}

\def\gapp{\ \lower 3pt\hbox{${\buildrel > \over \sim}$}\ }
\def\lapp{\ \lower 3pt\hbox{${\buildrel < \over \sim}$}\ }

\newcommand{\ben}{\begin{enumerate}}
\newcommand{\een}{\end{enumerate}}

\newcommand{\kms}{\,\,{\rm km\,\,s}^{-1}}

\newcommand{\be}{\begin{equation}}
\newcommand{\ee}{\end{equation}}
\newcommand{\bea}{\begin{eqnarray}}
\newcommand{\eea}{\end{eqnarray}}
\newcommand{\bean}{\begin{eqnarray*}}
\newcommand{\eean}{\end{eqnarray*}}
\newcommand{\rn}[1]{(\ref{#1})}

\newcommand{\next}{\nonumber \\}

\shorttitle{Extra-solar planetary systems}
\shortauthors{Dobbs-Dixon \& Lin \& Mardling}

\begin{document}

\title{Spin-Orbit Evolution of Short Period Planets} 

\author{Ian Dobbs-Dixon$^{1}$, D.N.C. Lin$^{1,2}$, and Rosemary A. Mardling
$^{1,3}$}

\affil{1) Department of Astronomy \& Astrophysics, 
University of California, Santa Cruz, CA 95064, USA} 
\affil{2) Institute of Astronomy, Cambridge University, Cambridge CB3 0HA, UK}
\affil{3) School of Mathematical Sciences, Monash University, 
Melbourne 3800, Australia}
\email{iandd@ucolick.org, lin@ucolick.org, rosemary.mardling@sci.monash.edu.au}

\begin{abstract}
The negligible eccentricity of all extra solar planets with periods
less than six days can be accounted for by dissipation of tidal
disturbances within their envelopes which are induced by their host
stars.  In the period range of 7-21 days, planets with circular orbits
coexist with planets with eccentric orbits.  These will be referred to
as the {\it borderline planets}.  We propose that this discrepancy can
be attributed to the variation in spin-down rates of young stars.  In
particular, prior to spin-down, dissipation of a planet's tidal
disturbance within the envelope of a sufficiently rapidly spinning
star can excite eccentricity growth, and for a more slowly spinning
star, at least reduce the eccentricity damping rate.  In contrast,
tidal dissipation within the envelope of a slowly spinning low-mass
mature star can enhance the eccentricity damping process.  Based on
these results, we suggest that short-period planets around relatively
young stars may have a much larger dispersion in eccentricity than
those around mature stars.  We also suggest that because the rate of
angular momentum loss from G and K dwarfs via stellar winds is much
faster than the tidal transfer of angular momentum between themselves
and their very-short (3-4 days) period planets, they cannot establish
a dynamical configuration in which the stellar and planetary spins are
approximately parallel and synchronous with the orbital frequency.  In
principle, however, such configurations may be established for planets
(around G and K dwarfs) with orbital periods of up to several weeks.
In contrast to G and K dwarfs, the angular momentum loss due to
stellar winds is much weaker in F dwarfs. It is therefore possible for
synchronized short-period planets to exist around such stars. The
planet around Tau Boo is one such example.
\end{abstract}

\section{Introduction}
Radial velocity surveys around nearby solar-type stars have led to the
discovery of planets with short-period and highly eccentric orbits
(Marcy {\it et al.} 2000). The period distribution is essentially
logarithmic ranging from a few days to many years.  While the upper
range of the period distribution at this stage is incomplete due to
relatively short survey times, there is a minimum cutoff at around
period $P=3$ days.  The observed eccentricities, $e$, of planets with
$P>21$ days are uniformly distributed in the range zero to about 0.7.
However, all planets with $P < 6$ days have nearly circular orbits
(Figure~\ref{figure1}(a)).  In this paper, we consider the effect of 
star-planet tidal interactions on the eccentricity
distribution of the short-period planets (hot Jupiters).

The dichotomy in the planets' eccentricity distribution is similar to
that of binary stars (Mathieu 1994).  It is generally assumed that
the orbital eccentricities of binary stars with periods less than some
critical value (the circularization period $P_c$) are damped by their
tidal interaction (Zahn 1977).  The magnitude of $P_c$ is a function
of the tidal dissipation efficiency and the age of the
binary.  The former quantity is usually parameterized in terms of the stars'
$Q_*$-values (cf Murray \& Dermott 1999; in what follows, an asterisk subscript will refer
to stellar values while the subscript $p$ will refer to planets).  Based on the
observed magnitude of $P_c$ in various stellar clusters (Mathieu
1994), the inferred values of $Q_*^\prime = 3 Q/2k_*$ (where $k_*$ is the tidal Love 
number or twice the apsidal motion constant) are $\sim 1.5
\times 10^5$ (with $P_c \sim 4$ days) for young stars with age $\tau_\ast
<10^8$ yr, and $\sim 10^6$ for mature stars (Terquem {\it et
al.}  1998).

The circularization of short-period planetary orbits may also be due
to tidal dissipation processes (Rasio {\it et al.}
1996, Marcy {\it et al.} 1997).  For example, this would be necessary to
account for the small eccentricities of short-period planets if they
were scattered into the vicinity of
their host stars by other planets (Rasio \& Ford 1996, Murray {\it et al.} 1998).  
In contrast, if short-period
planets acquired their small semi-major axes through tidal interaction
with their nascent disks (Lin {\it et al.}  1996), planet-disk
resonant interactions would naturally damp the eccentricities of
planets with masses $M_p < 10 M_J$ (where $M_J$ is the mass of
Jupiter) (Goldreich \& Tremaine 1980, Artymowicz 1992, Papaloizou {\it
et al.}  2001) although in some circumstances, it may also excite
planets' eccentricities (Goldreich \& Sari 2003).  The processes
(planet-star tidal interaction, unipolar induction, and disk
depletion) which halted the orbital migration of hot Jupiters (Lin
{\it et al.}  1996) may also induce the excitation of orbital
eccentricity. In either early dynamical evolution scenarios, efficient
post-formation eccentricity damping is needed to account for the small
eccentricities ($< 0.1$) of the short-period planets.

The orbital semi-major axes of short-period planets are larger than the
radii of their host stars by an order of magnitude and of their own
radii by two orders of magnitude. The angular momenta of these
planetary orbits are comparable to that carried by the spin of their host
stars and larger than that contained in their own spins.  Unless their
orbits are synchronized and circularized, these planets are subject to
intense tidal disturbances from their host stars.  
The $Q_p$-value for Jupiter has been inferred from
Io's orbital evolution such that $5 \times 10^4 < Q_p^\prime < 2
\times 10^6$ (Yoder and Peale 1981). With this range for $Q_p^\prime$,
planets with periods less than 6 days whose masses and radii are
comparable to those of Jupiter are circularized during the main
sequence lifetime ($\tau_\ast \sim$ a few Gyr) of solar-type dwarf
stars.  This estimate is consistent with the observed circular orbits
of the shortest period planets (Trilling 2000).

The dissipation of these disturbances leads not only to dynamical
changes but also to internal heating which may cause them to inflate
(Bodenheimer {\it et al.}  2001).  With a sufficiently large
eccentricity, hot Jupiters would undergo runaway tidal inflation and
overfill their Roche lobes before their orbits became circularized (Gu
{\it et al.}  2003, 2004).  Mass loss through Roche lobe overflow may
either totally disrupt the planet or induce the orbital semi-major
axis, $a$, to expand (Trilling {\it et al.} 1998). Both outcomes
inhibit protoplanets from becoming too close to their host stars.
Thus, the excitation and damping of eccentricity also has
observable consequences and important implications for the structure
and survival of individual planets and the period distribution of
planetary systems.

The observed eccentricity distribution also indicates that planets
with periods between 1 and 3 weeks have non-negligible eccentricities
which are smaller than those of planets with longer periods and larger
than those of planets with shorter periods.  We will refer to planets
in this period range as {\it borderline planets}.  In this paper, we
adopt a conventional approach using
frequency independent $Q_{p, \ast}^\prime$-values
to analyze the orbital evolution of a star-planet system (Mardling \& Lin 2002). 

In \S2, we show that the observationally inferred eccentricities of
these borderline planets are not well correlated with the expected
time scale, $\tau_{ep}$, for eccentricity damping due to the
dissipation of the stellar tidal disturbance inside the planets.
Although it is customary to assume that dissipation inside a planet
provides the dominant contribution to its orbital eccentricity
evolution, dissipation within its host star can have a comparable effect if
$Q_\ast^\prime \sim Q_p ^\prime$. We examine the possibility that, in
this limit, the dispersion in the ($e - \tau_{ep}$) relationship is
caused by the observed spread in stellar rotation rates.  Stellar spin
evolves as a consequence of structural changes within a star, as well
as mass loss and tidal interaction with a planet.

In \S3, we consider how these effects combined affect the long term
evolution of a star's spin as well as the eccentricity of a planetary
orbit.  We show that short-period planets rapidly evolve into a
spin-orbit quasi-equilibrium which continues to evolve, albeit at a
slower pace, because the dissipation of the stellar tidal disturbance
within a planet continues to induce eccentricity damping.  In
contrast, during the early stages of its main sequence evolution, a
solar-type star rotates rapidly and is likely to excite the
eccentricity of its planet.  However, as it loses angular momentum and
attains a modest or slow spin frequency, the tidal dissipation within
such a star will induce eccentricity damping.  We also deduce the
condition for quasi-equilibrium in which the loss of angular momentum
through a stellar wind is balanced by that transferred to a star
through its tidal interaction with a short-period planet.  We then
explore the possibility of such a system evolving into a state in
which the stellar and planetary spins are aligned, and the planet and
the envelope of the star spin synchronously with the orbit.  Finally
in \S4 we summarize our results and discuss their implications for the
dynamical evolution of short-period planets and the spin-down history
of their host stars.

\section{The Eccentric Short-Period Planets}

Observational determinations of planetary eccentricities require
extremely accurate radial velocity data over many orbits.  Spurious
determinations of periods and eccentricities may be caused by sparse
sampling, modest velocity measurement errors, and perturbations by
other planets.  Neglecting these potential sources of error in the
published data, we find that half of the known systems with $1$ week
$< P< 3$ weeks have non negligible eccentricities.  (We consider the
present threshold of real detection to be $ e = 0.1$ which corresponds
to the typical ratio of residual to amplitude of radial velocity
curves.)  In this section, we consider the implications of these
measurements.

\subsection{Dispersion in the Eccentricity Evolution of Short-Period Planets}
Using the standard formula (Goldreich \& Soter 1966, Yoder \& Peale
1981, Peale 1999, Murray \& Dermott 1999, for the limitation and
uncertainties see \S 1), the timescale associated with eccentricity
damping due to dissipation of the stellar tidal disturbance in a
planet can be expressed as
\begin{equation}
\tau_{ep} = -{e \over \dot e} = {4 \over 63} {M_p  \over M_\ast}
\left({a \over R_p} \right)^5 {Q_p^\prime \over n}
\simeq 5 \left({Q_p^\prime \over 10^6} \right)
\left({M_p \over M_J} \right) \left({M_\ast \over M_\odot}
\right)^{2/3} \left( { P \over {\rm 1\, day} }\right)^{13/3}
\left({R_p \over R_J} \right)^{-5} {\rm Myr},
\label{eq:taue}
\end{equation}
where $M_\ast$ is the mass of the star, $a$ is the semi-major axis,
and $n$ is the orbital mean motion.  The derivation of this formula
assumes $e\ll 1$ as well as aligned and synchronous planetary
rotation, that is, $\Omega_p=n=(GM_\ast /a^3)^{1/2}$.
We will use it as a reference timescale to illustrate the effects of
significant eccentricity and stellar spin.

In Figure~\ref{figure1}(b) we plot eccentricity as a function of
$\tau_{ep}$ of all known planets with $P<2$ months.  The period range
in this sample is selected to avoid the observational incompleteness
for planets with longer periods.  The magnitude of $Q_p ^\prime$ is
assumed to be $10^6$ (see discussions in \S2.3).  The results in
Figure~\ref{figure1}(b) clearly show that $e$ is negligibly small
($<0.1$) for planets with $\tau_{ep} < 1 $ Gyr.  In the transition
range where 1 Gyr $<\tau_{e p} < 10$ Gyr, planets on circular orbits
($e < 0.1$) coexist with those on eccentric orbits ($e> 0.1$).  This
dispersion cannot simply be attributed to the difference in the
stellar ages ($\tau_\ast$) since the target stars are chosen for their
lack of chromospheric activity which generally indicates $\tau_\ast
\sim$ a few Gyr (see Figure~\ref{figure1}(c) for $e$ versus $\tau_{e
p} / \tau_\ast$ ).  In addition, their dynamical properties cannot be
the main cause because some planets have quite different
eccentricities despite the similarity between their periods and mass
ratios. Although modest variations in $R_p$ may significantly modify
the magnitude of $\tau_{e p}$, planets with $P > 3-4$ days cannot be
easily inflated by tidally dissipation before their eccentricities are
damped out (Gu {\it et al.}  2003). The insolent effect of stellar
irradiation in quenching a planet's Kelvin-Helmholtz contraction
(Burrows {\it et al.} 2000) is also not well correlated with
eccentricity (see Figure~\ref{figure1}(d) for $e$ versus stellar flux
at $a$).  If the short-period planets were scattered to their present
location through dynamical instabilities on a variety of timescales,
large eccentricities ($\sim 1$) would be expected for all planets with
$\tau_{e p} > \tau_\ast$ and some planets with $\tau_{e p}$ smaller
than $\tau_\ast$.  This inference is not consistent with the results
shown in Figure~\ref{figure1}(c).  We cannot yet rule out the
possibility that the dispersion in the $e-\tau_{ep}$ relation is
due to some large-amplitude fluctuations in the $Q_p^\prime$ value
which itself may be determined by some stochastic processes (see \S2.3).

\subsection{Dissipation in the Host Star of the Planetary Tidal 
Disturbance}\label{planets}
The expression in eq(\ref{eq:taue}) does not include the contribution
from the dissipation in the star of the planet's tidal disturbance.
In the case that both the stellar and planetary spins are aligned with
the orbit, the rate of change of $e$ (Eggleton {\it et al.} 1998,
Mardling \& Lin 2002) becomes
\begin{equation}
\dot e = g_p + g_\ast
\label{eq:edot}
\end{equation}
where
\begin{equation}
g_{p,\ast} = \left( \frac{81}{2}\frac{n\,e}{Q_{p,\ast}^\prime}\right)
\left({M_{\ast, p} \over M_{p, \ast}} \right) 
\left( {R_{p, \ast} \over a} \right)^5 \left[ -f_1 (e) 
+ \ff{11}{18} f_2 (e)\left( \frac{\Omega_{p, \ast}}{n}\right)\right], 
\label{eq:edotg}
\end{equation}
\begin{equation}
f_1(e) = \left(1 + \ff{15}{4} e^2 + \ff{15}{8} e^4 + \ff{5}{64} e^6 
\right)/ (1 - e^2 )^{13/2},
\end{equation}
\begin{equation}
f_2(e) = \left(1 + \ff{3}{2} e^2 + \ff{1}{8} e^4 \right)/
(1 - e^2 )^{5},
\end{equation}
For $e\ll1$,
\begin{equation}
g_{p,\ast} = \left( \frac{81}{2}\frac{n\,e}{Q_{p,\ast}^\prime}\right)
\left({M_{\ast, p} \over M_{p, \ast}} \right) 
\left( {R_{p, \ast} \over a} \right)^5 
\left[-1+\ff{11}{18}\left(\frac{\Omega_{p,\ast}}{n}\right)\right].
\label{eq:gpast}
\end{equation}
For the case in which only a synchronously spinning
planet contributes to $\dot e$, eqs (\ref{eq:edot}) and
(\ref{eq:gpast}) reduce to eq(\ref{eq:taue}).

Equations \rn{eq:edot} and \rn{eq:edotg} allow us to examine the
orbital evolution of star-planet systems with non-synchronized spins
and arbitrary eccentricities.  For example, a system in which the
stellar tide can be neglected, (for instance, if the star were
extremely compact), the tidal dissipation in the planet leads to
eccentricity excitation if
\begin{equation}
\Omega_p > \Omega_{p c} \equiv\ff{18}{11} f_1(e) n / f_2(e)\rightarrow \ff{18}{11}n
\end{equation}
as $e\rightarrow 0$ (also see Goldreich \& Soter 1966).  We now
introduce two parameters which characterize the influence of the
stellar tide and spin, as well as arbitrary eccentricity, on the
orbital evolution of a system. The {\it eccentricity damping
efficiency factor} $\beta$ is defined as
$\beta\equiv{\tau_{ep}}/{\tau_{e}}$, where \be
\tau_e=-\frac{e}{\dot{e}}=\frac{\tau_{ep}}{\beta}
\label{eq:taueb}
\ee is the eccentricity damping timescale. For $\beta<1$, stellar
tides inhibit eccentricity damping, whereas for $\beta>1$ they
enhance the damping rate.  Negative values of $\beta$ corresponds to
eccentricity excitation and increasing the eccentricity always
enhances the damping rate.  From equations~\rn{eq:taue} to
\rn{eq:edotg} we have \be \beta = {18 \over 7} \left\{\left[ f_1 (e) -
\ff{11}{18}f_2(e)\left(\frac{\Omega_p}{n}\right)\right]+ \left[
f_1(e)-\ff{11}{18}f_2(e)\left(\frac{\Omega_\ast}{n}\right)
\right]/\lambda\right\},
\label{eq:beta}
\end{equation}
where
\be
\lambda=\left( {Q_\ast ^\prime \over Q_p^\prime} \right) \left( {M_\ast \over 
M_p } \right) ^2 \left( {R_p \over R_\ast} \right)^5.
\label{eq:lambda}
\ee Here we have introduced a second parameter, $\lambda$, to
characterize the importance of the relative masses, radii and
$Q$-values of the star and planet.  Generally, tidal dissipation in
the planet dominates the evolution when $\lambda$ is large.

\subsection{Magnitude of the Stellar and Planetary Q-values}

In eq (\ref{eq:lambda}), the value of $M_p/M_\ast$ can be directly
determined from observational data (to within an uncertainty of $\sin
i$, where $i$ is the inclination of the system to the line of sight).
The value of $R_\ast/R_p$ can be deduced theoretically from stellar
and planetary structure models or inferred empirically from the
observed radius of the transiting planet HD~209458. The major
uncertainty is associated with the magnitude of the $Q$-value because
the dominant physical processes which determine the magnitude of
$Q_{p,\ast}^\prime$ in both solar type stars and gas giant planets
remain unresolved.  Despite the availability of observationally
inferred $Q$-values, it is not clear whether those obtained for
solar-type binary stars and for Jupiter (see \S1) can be directly
applied to hot Jupiters.

Theoretical analyses of tidal evolution mainly focus on the response and dissipation of
both equilibrium and dynamical tides.  The equilibrium model is based
on the concept that a homogeneous spherical body continually adjusts
to maintain a state of quasi-hydrostatic equilibrium in the varying
gravitational potential of its orbital companion. Internal friction
within the body induces the dissipation of energy and a phase lag
which gives rise to a net tidal torque which transports angular
momentum between the spin of the body and its companion and their
orbits.  Although turbulence can lead to dissipation in the extended convective envelopes of gaseous
giant planets and low-mass stars,
the convective turnover timescale is usually much longer than the
period of the tidal forcing. Due to this reduction in the efficiency
of the turbulent viscosity (Zahn 1977, 1989, Goodman \& Oh 1999), the
derived rate of angular momentum transfer falls short of that required
for the circularization of solar-type binary stars by nearly two
orders of magnitude (Terquem et al. 1998). For gas giant planets,
turbulent dissipation of equilibrium tides gives $Q\approx
5\times10^{13}$ (Goldreich \& Nicholson 1977).

In radiative stars, the tidal perturbation of the companion can induce
the resonant excitation of low-frequency g-mode oscillations in the
radiative region, which carry both energy and angular momentum fluxes
(Cowling 1941; Zahn 1970). When this wavelike response is propagated
to the stellar surface, radiative damping of the dynamical response
provides an effective avenue for tidal dissipation in high-mass binary
stars (Savonije \& Papaloizou 1983, 1984). However, the envelope of
Jupiter is mostly convective (Guillot et al. 2003) and the relevance of
g-mode oscillations is less well established (Ioannou \& Lindzen
1993a,b, 1994).  Nevertheless, the surface layers of hot Jupiters may
attain a radiative state because they are intensely heated by their
host stars. If so, g-mode oscillations may be excited just above the
planet's convective envelope and dissipated through radiative or
nonlinear damping (Lubow, Tout, \& Livio 1997) as suggested for
high-mass stars (Goldreich \& Nicholson 1989). In rotating gaseous
planets, Coriolis force can provide a restoring force to the tidal
perturbation and induce the resonant excitation of a rich spectrum of
inertial modes (Savonije et al. 1995).  When the forcing and
initial-wave frequencies resonate with each other, the tidal response
of the star is greatly enhanced (Papaloizou \& Savonije 1997).  Hot
Jupiters probably formed as rapid rotators such that their tidal
forcing occurs in the frequency range of inertial waves, which
therefore constitute the natural and dominant response of the planet
(Ogilvie \& Lin 2004).

In principle, the value of $Q_{p,\ast}^\prime$ depends in a highly
erratic way on the resonant response which is a function of both the
planet's spin and the perturber's forcing frequencies as well as the
magnitude of viscosity.  In the presence of viscosity, the
spectrum of inertial waves is discrete, but as the viscosity 
tends to zero, the spectrum
becomes increasingly dense (e.g., Dintrans \& Ouyed 2001). In this limit,
the dissipation rate may not simply vanish because of the increasing
probability of resonant excitation of wave modes.
Although the highly variable $Q_p$ values can
introduce a dispersion in the rate of eccentricity dissipation, the
spin of a planet changes as it undergo Kelvin Helmholtz contraction
such that the relevant frequency-averaged $Q$-value ($\sim 10^6$) may
be asymptotically independent of the viscosity in the limit of small
Ekman number (Ogilvie \& Lin 2004).

Based on these considerations and for computational convenience, we
adopt the equilibrium-tide prescription in eq(\ref{eq:edot}) for the
determination of the eccentricity evolution of hot Jupiters.

\subsection{Eccentricity Evolution of Planets around Rapidly Spinning Stars}

The stellar tide can become stronger than the planetary tide if the spin
rate of the star is high enough. From eqns\rn{eq:edot} and \rn{eq:edotg} the
critical stellar spin rate corresponding to $\dot e=0$ is
\be
\Omega_{\ast c}\equiv\frac{18}{11}\left[\left(1+\lambda\right)\frac{f_1(e)}{f_2(e)}
\right]n-\lambda\,\Omega_p,
\label{eq:omeast}
\ee
so that the eccentricity grows if $\Omega_\ast>\Omega_{\ast c}$.
If $\Omega_p \simeq n$ and $e\ll1$,  
\be
\Omega_{\ast c}=(18+7\lambda)n/11.
\label{eq:omeast1}
\end{equation}
For Jupiter-like planets orbiting solar-like stars, $\lambda=\lambda_{J\odot}\simeq 12$
so that $\Omega_{\ast c}\simeq 9n$.

In the next section, we show that planets attain a state of
synchronization ($\Omega_p=n$) on a timescale ($\tau_{\Omega_p}$)
which is generally much shorter than that of their host stars
($\tau_{\Omega \ast}$). We therefore now consider systems in which the
planet is nearly synchronously rotating.  In Figure~\ref{beta}(a) we
plot $\beta$ (eqn\rn{eq:beta}) as a function of stellar spin period
for various values of $e$.  For illustration purposes we use
$\lambda=\lambda_{J\odot}$ with $Q_p \simeq Q_\ast$. In addition, we
have taken the orbital period to be 6.5 days, representing the three
shortest period borderline planets.  Now recall that $\beta>1$
corresponds to enhanced eccentricity damping, $0<\beta<1$ corresponds
to prolonged circularization times, while $\beta<0$ corresponds to
eccentricity excitation.  Also note that stars with radii the same as
the Sun's and rotating with velocities $V_r= 50\kms$ have spin periods
of 1 day, while those with $V_r=10\kms$ have spin periods of 5 days
(see next section).  For each value of $\Omega_\ast$, $\beta$ is
bounded below by the value corresponding to $e=0$ (dotted curve).  For
high values of $e$, $\beta>1$ for most stellar spin rates, suggesting
that eccentric borderline planets cannot have reached their present
positions with eccentricities much higher than presently observed.

Figure~\ref{beta}(b) plots $\beta$ against $\Omega_\ast$ for the three
shortest period borderline systems HD~\-168746 ($e=0.08$,
$M_\ast=0.92M_\odot$, $M_p\sin i=0.23M_J$, orbital period $P=6.4$
days, $\tau_{ep}=0.60-6.0$ Gyr, $\lambda=127.4$; Pepe et al.\ 2002),
HD~217107 ($e=0.14$, $M_\ast=0.98M_\odot$, $M_p\sin i=1.30M_J$,
$P=7.1$ days, $\tau_{ep}=1.90-19.0$ Gyr, $\lambda=7.2$; Fischer et
al.\ 1999), and HD~68988 ($e=0.15$, $M_\ast=1.2M_\odot$, $M_p\sin
i=1.90 M_J$, $P=6.3$ days, $\tau_{ep}=1.77-17.7$ Gyr, $\lambda=5.0$;
Vogt et al.\ 2002).  The ranges for $\tau_{ep}$ are calculated using
$Q_\ast=Q_p=10^5-10^6$ to represent the evolution of these parameters.
The orbital evolution of HD~168746 was clearly rapid, since any
eccentricity greater than its present value would enhance the damping
rate even more (see Figure~\ref{beta}(b)).  It is likely that the
present value is, in fact, much closer to zero.  Dissipation within
the planet dominated the evolution because the planet-to-star mass
ratio is relatively small, and the radii of Saturn-like planets are
relatively large (Burrows {\it et al.} 1997). Taking $R_p=R_{\rm
Saturn}=0.84R_J$ (and $\sin i=1$) gives the value for $\lambda$ quoted
above In contrast, the eccentricity evolution of both HD~217107 and
HD~68988 would have been retarded if the stars were rapid rotators
initially.

\subsection{Dispersion in the Stellar Spin Frequency}
The spin frequency of a star is generally believed to be a function of
its age, $\tau_\ast$ (see \S3).  However, in open clusters where
member stars presumably have very similar ages, an order of magnitude
dispersion has been observed in the stellar rotational velocity. For
example, although most G dwarfs in the Pleiades are observed to have
rotational speed $V_r \sin i = \Omega_\ast R_\ast \sin i< 10$ km
s$^{-1}$, at least $20 \%$ of them are ultra-fast rotators (UFRs) with
$30\kms<V_r\sin i < 100\kms$ (Soderblom {\it et al.} 1993a,b).  The
spin frequencies of these UFRs exceed the mean motion of planets with
periods of several days. (For a star with a solar radius, for example,
$V_r = 30$ km s$^{-1}$ corresponding to a spin period $2 \pi/
\Omega_\ast = 1.7$ days.)  In order to illustrate the importance of
stellar spin, we carried out a Monte Carlo simulation in which a
sample of planetary candidates was generated with the observed
logarithmic period distribution between 3 days and 3 years.  By
adopting $M_p = 1M_J$, $R_p =1.35 R_J$, $M_\ast=1 M_\odot$, $R_\ast=1
R_\odot$, $Q_p ^\prime = 10^6$, and $Q_\ast ^\prime = 1.5 \times
10^5$, and taking the planetary spins to be synchronous with the
orbital motion, we deduced values for $\beta$ and $\tau_e$ given the
observed spin periods of Pleiades stars and the simulated mean motions
$n=2\pi/P$.  In Figure~\ref{figure3}(a), we plot the inferred values
of $\tau_e/\tau_\ast$ (the age of the Pleiades cluster is taken to be
$7 \times 10^7$ yr) for various simulated values of $P$.  The open
circles represent eccentricity damping timescales while the filled
dots represent eccentricity excitation timescales. For all planets
with $P<5$ days, the eccentricity is damped on timescales shorter than
$\tau_\ast$.  For borderline planets with 5 days $ < P <$ 15 days, the
eccentricities of several systems are damped on timescales $\tau_e >
\tau_\ast$.  However, eccentricity is excited for several other
systems in this period range.  For $P > 15$ days, the eccentricities
of most systems are excited on timescales $\tau_e \gg \tau_\ast$.

Ultra-fast rotators with $V_r \sin i$ up to 200 km s$^{-1}$ have been
observed among the stars of $\alpha$ Persei (Prosser 1992), an open
cluster which is younger ($\tau_\ast \simeq 5 \times 10^7$ yr) than
the Pleiades. Although the time available for eccentricity evolution
is limited, the relatively small $Q_\ast^\prime$ values ($\sim 1.5
\times 10^5$ as inferred from observations of binary stars in these
young clusters) together with medium to high stellar spin rates imply
an enhanced role for the stellar tide and hence retarded eccentricity
damping or even eccentricity excitation in some cases.  Thus in these
young clusters, we expect planets with periods greater than 1-2 days
(including those with $\tau_{ep } < \tau_\ast$) to have a noticeable
dispersion in their period-eccentricity distribution.

In older clusters such as the Hyades for which $\tau_\ast \sim 6
\times 10^8$ yr, $V_r \sin i$ is below $10 \kms$ for nearly all G and
K dwarfs.  In contrast, most F dwarfs are still fairly rapidly
spinning, with rotation speeds in the range $20-70\kms$.  We performed
a second Monte Carlo simulation, this time for the Hyades, adopting
$Q_\ast =10^6$ and $\tau_\ast = 6 \times 10^8$ yr. We found
(Figure~\ref{figure3}(b)) that the majority of the G and K dwarfs have
slowed down sufficiently to promote eccentricity damping of
short-period planets.  There are nevertheless a few rapidly rotating F
dwarfs which continue to promote eccentricity excitation.

\subsection{Stellar and Planetary Structural Adjustments}\label{2.4}
Eqn (\ref{eq:edot}) indicates that $\dot e$ is a rapidly increasing
function of $R_p$.  The evolution of $a$ and $\Omega_{p,\ast}$ also
depend sensitively on $R_p$ and $R_\ast$ (see the next section).
During the early stages of stellar evolution, both stellar and
planetary radii are somewhat larger than their values at later stages.
We now assess the effects of stellar and planetary structural
adjustment in regulating eccentricity and semi-major axis evolution.
For computational convenience, we approximate existing numerical
stellar models (Iben 1965, 1967) with the analytic formula
\begin{equation}
R_\ast \simeq R_\odot (\tau_\ast / 10^7 {\rm yr})^{-0.3}
\end{equation} 
for $\tau_\ast < 10^7$ yr.  This timescale for solar-type stars to
undergo Kelvin Helmholtz contraction onto the main sequence is much
shorter than $\tau_e$ or $\tau_{\Omega \ast}$, such that the initial
large stellar radii cannot significantly modify $e$ or $\Omega_p$.

However, planets have much slower Kelvin-Helmholtz contraction rates,
(Burrows {\it et al.} 2000, Bodenheimer {\it et al.} 2000), and are
such that
\begin{equation}
R_p \simeq 1.2 R_J (\tau_\ast / \tau_o )^{-0.1},
\label{eq:rp}
\end{equation} 
where $\tau_o=1$ Gyr.  For $\tau_\ast=\tau_o$, $R_p$ in this
prescription is actually smaller than that used in our models above
($R_p=1.35 R_J$).  The larger value takes into account stellar
insolation of a short-period planet's surface, a process which tends
to extend its contraction timescale (Burrows {\it et al}. 2000) and
tidal inflation (Bodenheimer {\it et al}. 2001).  Using
eqs(\ref{eq:edot}) and (\ref{eq:rp}) for an unheated planet, we find
that $\tau_e$ is an order of magnitude smaller when $\tau_\ast=10^8$
yr than when $\tau_\ast=4.6$ Gyr. (For $\tau_\ast = 10^{7.5-8.5}$ yr,
a planet's size is essentially independent of its mass.)
Nevertheless, during the first $10^8$~yr, $\tau_e/\tau_\ast >1$ and
there is insufficient time to circularize the orbits of borderline
planets.  However, since $\tau_{e p} / \tau_\ast \propto \tau_\ast
^{-0.5}$, eccentricity damping occurs mostly during the later stages
of main sequence evolution and borderline planets can be circularized
provided their host stars are not rapid rotators.

In addition to Kelvin Helmholtz contraction, a planet's size is
modified by tidal dissipation which leads to an internal energy
generation rate
\begin{equation}
\dot E_t = {G M_\ast M_p e \vert \dot e\vert _p \over a (1 - e^2)} + 
\alpha_p M_p R_p^2\, \Omega_p \vert \dot \Omega_p \vert
\end{equation}
where $\vert \dot e \vert_p$ refers to the planet's contribution to
the circularization process in eq(\ref{eq:edot}) and $\dot \Omega_p$
is the rate of change of $\Omega_p$ (see next section).  For
sufficiently large $\vert \dot e \vert_p$ or $\vert \dot \Omega_p
\vert$, the magnitude of $\dot E_t$ exceeds the rate of energy
released $\dot E_{KH} (M_p, R_p)$ due to an unheated planet's
Kelvin-Helmholtz contraction.  In this limit, a planet's radius
evolves toward an equilibrium value $R_e$ such that its surface
luminosity, ${\cal L}(M_p, R_e)$, is balanced by $ \dot E_t$
(Bodenheimer {\it et al.} 2001).

However, the evolution of $R_p$ in eq(\ref{eq:rp}) depends on $\dot
E_{KH}$.  If a short-period planet arrives at its current location
shortly after formation, the value of $\dot E_{KH}$ at the time will
exceed $\dot E_t$. After that, Kelvin-Helmholtz contraction of the
planet will stall once $\dot E_{KH}$ becomes less than $\dot E_t$. On
the other hand, if a short-period planet acquires its present orbital
properties {\it after} it contracts to a radius $R_p<R_e$, its size
will adjust at a rate (Mardling \& Lin 2002)
\begin{equation}
\dot R_p \simeq {G M_p^2 \over R_p \dot E_t},
\end{equation}
with an enlargement timescale 
\begin{equation}
\tau_R = {R_p \over \dot R_p}
\simeq {M_\ast R_p \over M_p a} {e^2 \over (1 - e^2)} \tau_{e p}. 
\end{equation}
This timescale is generally longer than $\tau_{\Omega p}$, in which
case a planet cannot inflate significantly before it attains a state
of near synchronization (Gu {\it et al.} 2003).

But, interior to $a \sim 0.04  $AU, $\tau_R < \tau_{e p}$ for a planet
with  non-negligible  eccentricity ($>0.1$)  so  that  it can  inflate
before the  eccentricity is damped out.   Planetary inflation enhances
the rate of  tidal dissipation and shortens $\tau_e$.  For some values
of  $e$  and  $a$, $R_e$  exceeds  a  planet's  Roche radius,  $R_R  =
(M_p/3M_\ast)^{1/3} a$, so  that it begins to lose  mass through Roche
lobe overflow while its radius is  confined to $R_p = R_R$ (Gu {\it et
al.}  2003).   Exterior  to  $0.04$  AU, the  condition  required  for
planetary  inflation ($\tau_R  <  \tau_{e p}$)  is  more difficult  to
satisfy unless the actual eccentricity-damping timescale, $\tau_e$, is
significantly  prolonged by eccentricity-excitation  due to  the tidal
dissipation inside a rapidly rotating host star, or due to the secular
perturbation of  other planets. Nevertheless, $\dot E_t$  is a rapidly
decreasing function of  $a$.  Thus for a borderline  planet, if $a$ is
sufficiently  large  and  $\dot  E_t$  sufficiently  small,  $R_e$  is
unlikely to be much larger than our adopted values of $1.35-1.5 R_J$.

\subsection{Eccentricity Dispersion Among Borderline Planets}

Among the known planets, the shortest period planet with a significant
eccentricity is HD~68988, with $e=0.15$, $P=6.276$ days, and minimum
mass $M_p \sin i = 1.9 M_J$.  The mass and age of the host star are
$M_\ast = 1.2\, M_\odot$ and $\tau_\ast \simeq 6$ Gyr respectively.
Given these values and taking $R_p=1.35 R_J$, $\tau_{e p} \simeq
\tau_\ast$ if $Q_p ^\prime \simeq10^6$ (Vogt {\it et al.} 2002).  The
value for $R_p$ is that inferred from transit observations of the
short-period planet around HD~209458 (Brown {\it et al.} 2001).

The dynamical properties of HD~68988 are remarkably similar to those
of HD~217107 which has a similar eccentricity ($e = 0.14$).  However,
there are several other planets such as HD~168746 and HD~130322 which
have similar values for $\tau_{e p}$ but negligible eccentricities. We
propose that the dispersion in eccentricity of borderline planets is
correlated with the dispersion of stellar spin frequencies of young
stars.  Based on the lack of chromospheric activity, Vogt {\it et al.}
(2002) inferred that HD~68988 is a slow rotator.  If correct, the
stellar tide will be promoting eccentricity damping in this system
today.  Using the Skumanich law (see eqn~\rn{eq:vrot} below) as well
as data from Soderblom {\it et al.} (1993a, b), we find that the mean
spin frequency of F, G, and K stars with an age comparable to that of
the Pleiades cluster is about half the present orbital mean motion of
the planet around HD~68988, {\it i.e.}  $\Omega_\ast < \Omega_{p c}$.
However, as discussed above, slow and fast rotators coexist in this
cluster, with the dispersion in the observed values of $V_r\sin i$
being large.  There also exist solar-type stars which are older than
the Sun but which are more rapidly spinning.  If HD~68988 and
HD~217107 were rapid rotators when they were young, at that time
$\beta<1$ so that $\tau_e > \tau_{e p}$ (Figure~\ref{beta}(b)).  The
effect of a young star's tide is also enhanced by its slightly smaller
value of $Q_\ast ^\prime$ as inferred from observation.  Similarly, if
HD~168746 and HD~130322 were slow rotators with $\Omega_\ast <
\Omega_{* c}$ since birth, $\tau_e<\tau_{ep}$.

To support our conjecture, we carried out some numerical simulations
using a scheme which calculates the tidal, spin and dynamical
evolution of a multi-planetary system (Mard\-ling \& Lin, 2002).
Using the observed kinematic data for HD~68988 together with $R_p =
1.5 R_J$, $R_\ast = 1.3 R_\odot$, $Q_\ast^\prime \simeq Q_p^\prime
\simeq 10^6$, $\Omega_p = n$, and $\Omega_\ast = 2 \pi/(2 $ days), we
obtain $\tau_{e} \sim \tau_\ast \sim $ a few Gyr.  However, if the
spin frequency of the star is halved, $\tau_e $ is reduced by a
similar factor.  The magnitude of $\tau_e$ becomes less than 1 Gyr for
slowly rotating host stars with $\Omega_\ast < n$. These numerical
results are in general agreement with our conjecture.  A more
comprehensive numerical analysis is described at the end of the next
section.

\section{Evolution of the Stellar and Planetary Spins}
The above results clearly demonstrate the importance of the stellar
tide and spin in the eccentricity evolution of short-period
planets. It seems reasonable, therefore, that the dispersion in the
$e-\tau_{ep}$ relation is a result of the spread of spins among young
stars.  However, most of the target stars in the radial velocity
surveys are chosen for their low level chromospheric activities.
These type of stars, including several borderline systems with
measurable eccentricities, have slow rotation speeds like the Sun.

Nevertheless, some, if not all, host stars of extrasolar planets may
have been rapidly spinning in the past. It is therefore worthwhile
investigating the evolution of $\Omega_\ast$ and $\Omega_p$.  Assuming
an aligned rotator and that $M_\ast\gg M_p$, the rate of change of the
stellar spin frequency is given by (Mardling \& Lin 2002)
\begin{equation}
\dot\Omega_\ast=\frac{9}{2}\left(\frac{n^2}{\epsilon_\ast\alpha_\ast Q^\prime_\ast}\right)
\left(\frac{M_p}{M_\ast}\right)^2
\left(\frac{R_\ast}{a}\right)^3
\left[f_3(e)-f_4(e)\left(\frac{\Omega_\ast}{n}\right)\right]+\dot\omega_\ast,
\label{eq:omedot}
\end{equation}
where \bea f_3(e) &=& (1 + \ff{15}{2} e^2 + \ff{45}{8} e^4 +
\ff{5}{16} e^6)/ (1 - e^2)^6,\\ f_4(e) &=& (1 + 3 e^2 + \ff{3}{8} e^4
)/(1 - e^2)^{9/2}, \eea and $\dot \omega_\ast $ is the rate of change
of the stellar spin due to angular momentum loss via a stellar wind
(see the next subsection).  The quantity $\epsilon_{\ast}$ is the
stellar mass fraction participating in tidally induced and external
angular momentum exchange and $\alpha_\ast$ is the moment of inertia
coefficient of the star.  The rate of change of the planet's spin is
given by
\begin{equation}
\dot\Omega_p=\frac{9}{2}\left(\frac{n^2}{\epsilon_p\alpha_p Q^\prime_p}\right)
\left(\frac{M_\ast}{M_p}\right)
\left(\frac{R_p}{a}\right)^3
\left[f_3(e)-f_4(e)\left(\frac{\Omega_p}{n}\right)\right]+\dot\omega_p,
\label{eq:omepdot}
\end{equation}
where $\epsilon_p$ and $\alpha_p$ are the planetary counterparts of
$\epsilon_\ast$ and $\alpha_\ast$.  Note that $\dot\Omega_\ast$
depends more strongly on the mass ratio than does $\dot\Omega_p$.
Since gaseous planets have extensive convection zones which can be
fully mixed, $\epsilon_p \simeq 1$.  In contrast, solar-type stars
have shallow surface convection zones so that $\epsilon_\ast$ is a few
times $10^{-2}$ for G dwarfs, and an order of magnitude smaller/larger
for F/K dwarfs.

\subsection{Timescales for Spin and Semi-Major Axis Evolution}

The expression for the rate of change of the stellar spin (eq
(\ref{eq:omedot}) ) involves two contributions, as does its
counterpart for the rate of change of planetary spin. For
post-formation evolution we can neglect $\dot\omega_p$ for planets,
while $\dot\omega_\ast$ is associated with angular momentum loss via a
stellar wind.  The tidal transfer of angular momentum induces
$\Omega_p$ to evolve on a timescale
\begin{equation}
\tau_{\Omega_p} \equiv \left( \frac{n-\Omega_p}{\dot\Omega_p }\right)_{\dot \omega_p =0}=
\frac{7}{2}\alpha_p\left(\frac{R_p}{a}\right)^2\tau_{ep}\ll \tau_{ep}
\end{equation} 
for non-synchronously spinning planets (see eqns~\rn{eq:taue} and
\rn{eq:omedot} with $e\ll1$ and $M_\ast\gg M_p$).  Therefore the
planetary synchronization process is much more rapid than the orbital
circularization process.  However, as a consequence of this, the
planetary spin can only be {\it quasi}-synchronized with the orbit
while the eccentricity is non-zero.  From eqn~\rn{eq:omepdot}, this
value is $\Omega_p \equiv f_3 (e) n / f_4 (e)\rightarrow (1+6e^2)n$
for $e\ll 1$.

For conservative systems ($\dot\omega_\ast=0$), the stellar spin evolves
on a timescale
\be
\tau_{\Omega_\ast} \equiv 
\left( \frac{n-\Omega_\ast}{\dot\Omega_\ast}\right)_{\dot \omega_\ast =0}=
\frac{7}{2}\alpha_\ast\epsilon_\ast\lambda\left(\frac{M_\ast}{M_p}\right)
\left(\frac{R_\ast}{a}\right)^2\tau_{ep}
\end{equation} 
so that for most systems of interest (in the absence of angular
momentum loss) we have that $\tau_{\Omega_\ast}>\tau_{ep}$.  Thus we
can conclude that $\Omega_\ast$ does not change significantly prior to
the circularization of a planet's orbit (in the context of solid
satellites around gaseous giant planets, the relative values of
$\tau_{\Omega_p}$, $\tau_e$, and $\tau_{\Omega \ast}$ are even more
extreme because $Q^\prime$ values for satellites are much smaller than
those for such planets; Goldreich \& Soter 1966).  We conclude,
therefore, that given the slow stellar rotation rates observed today
in borderline systems, were they to have suffered no angular momentum
loss in the past, tidal dissipation inside the stars could only ever
have enhanced eccentricity damping.

Assuming that the spins of a planet and a star are aligned with the
orbit normal of such a system, the total angular momentum
perpendicular to the orbit is
\begin{equation}
J_{\rm total} (a, \Omega_\ast, e)= J_o (a, e)+ J_p (a)+ J_\ast (\Omega_\ast),
\label{eq:jtotal}
\end{equation}
where $J_o=M_p a^2 n\sqrt{1-e^2}$ is the orbital angular momentum.
The angular momenta of the planet and star are \be J_p=\alpha_p M_p
R_p^2\,\Omega_p \ee and \be J_\ast=\alpha_\ast\epsilon_\ast M_\ast
R_\ast^2\,\Omega_\ast \ee respectively, where we have assumed that the
whole planet participates dynamically whereas only a fraction
$\epsilon_\ast$ of the star is involved in angular momentum exchange.
The rate of change of the total angular momentum is \bea
\dot{J}_{\omega_\ast}&=&\dot{J}_o+\dot{J}_p+\dot{J}_\ast \next
&=&J_o\left(\frac{\dot{a}}{2a}-\frac{e\dot{e}}{1-e^2}\right)+ \alpha_p
M_p R_p^2\,\dot\Omega_p +\alpha_\ast\epsilon_\ast M_\ast
R_\ast^2\,\dot\Omega_\ast \eea where \be
\dot{J}_{\omega_\ast}=\alpha_\ast\epsilon_\ast M_\ast
R_\ast^2\dot\omega_\ast
\label{eq:jdo}
\ee is the rate at which the system loses angular momentum via a
stellar wind.  In a conserved system, we find from Eqns~\rn{eq:omedot}
and \rn{eq:omepdot}, 
\be
\frac{1}{2}\frac{\dot{a}}{a}=e\dot{e}-\frac{9}{2}
\left\{\frac{1}{Q^\prime_p}\left(\frac{M_\ast}{M_p}\right)
\left(\frac{R_p}{a}\right)^5(n-\Omega_p)+
\frac{1}{Q^\prime_\ast}\left(\frac{M_p}{M_\ast}\right)
\left(\frac{R_\ast}{a}\right)^5(n-\Omega_\ast)\right\} \ee to first
order in the eccentricity. From Eqns~\rn{eq:taue} and \rn{eq:taueb},
this can be written as \be
\frac{\dot{a}}{a}=-\left(\frac{2e^2}{\beta}-\sigma\right)/\tau_{ep}
\label{eq:adot}
\ee where \be
\sigma=\frac{4}{7\lambda}\left(\frac{\Omega_\ast}{n}-1\right) \ee in
the limit that $\Omega_p=n$.  For small to moderate eccentricities and
for most stellar spin periods of interest, as well as for a wide range
of star-planet masses and radii (characterized by $\lambda$),
$|\beta|\sim O(1)$ (Figure~\ref{beta}).  However, as illustrated by
the shortest period borderline planets described in
Section~\ref{planets}, $\lambda$ can vary widely being $\sim O(1)$ for
systems with relatively massive planets and lower-mass stars, to
several hundreds for systems with lower mass planets and higher mass
stars.  For systems in which $\Omega_\ast>n$, $\dot{a}>0$ as long as
$e^2<\sigma\beta/2$. On the other hand, $\dot{a}<0$ whenever
$\Omega_\ast<n$.

For Jupiter-mass planets around solar-type stars, $\vert \dot J_p
\vert \ll \vert \dot J_\ast \vert$ so that the stellar spin provides
the dominant source of angular momentum exchange with the planet's
orbit.  In the absence of a stellar wind, stable configurations
satisfy $J_o>3J_\ast$ (Hut 1980, Lin 1981), or equivalently,
\begin{equation}
a > a_c \equiv \left[{3\alpha_\ast \epsilon_\ast \over {\sqrt
(1 - e^2)}} \left(\frac{M_\ast }{M_p}\right)
\left(\frac{\Omega_\ast}{n}\right)\right]^{1/2}R_\ast,
\label{eq:acrit}
\end{equation}
and angular momentum exchange acts to reduce the difference between
$n$ and $\Omega_\ast$ until an equilibrium configuration is achieved.
The timescale for the semi-major axis to evolve is
\begin{equation}
\tau_a = {a \over |\dot a|}
=\frac{3}{2}\left(\frac{a}{a_c}\right)^2\left(\frac{\Omega_\ast}{|n-\Omega_\ast|}\right)
\tau_{\Omega_\ast}.
\ee

\subsection{Simultaneous Evolution of the Stellar Spin and Planetary Orbit}
Stars are known to lose spin angular momentum via stellar winds so
that $\dot\omega_\ast$ in Eqn~\rn{eq:omedot} is finite and negative.  The
youngest stars are observed to have rapid rotational velocities which
are sometimes as high as their breakup speeds (Stassun {\it et al.}
1999).  Although observations of G and K dwarfs in the $\alpha$ Persei
and Pleiades clusters (with $\tau_\ast < 10^8$ yr) indicate a large
dispersion in $V_r \sin i$ (with magnitudes ranging from a few to
100-200 km s$^{-1}$), the same type of stars in the Hyades (with
$\tau_\ast = 6 \times 10^8$ yr) have nearly uniform $V_r \sin i$ which
are less than $10 \kms$ (Soderblom {\it et al.}  1993a,b).  The
simplest description of the decline in the mean rotation velocity of
mature G and K dwarfs is given by Skumanich (1972) such that
\begin{equation}
<V_r \sin i > \simeq V_o (\tau_\ast / \tau_o)^{-0.5}
\label{eq:vrot}
\end{equation}
where $V_o \simeq 4$ km s$^{-1}$ and $\tau_o =1$ Gyr.  A more general
description is given by
\begin{equation}
\left( {V_o \over V_r(\tau_\ast) \sin i} \right)^2 - \left( {V_o \over
V_1 \sin i} \right)^2 = { \tau_\ast - \tau_1 \over \tau_o} 
\label{eq:vrot1}
\end{equation}
which more accurately models both the dispersion in $V_r$ among young
stars at some initial time $\tau_1$ and the homogeneous low $V_r$
among mature stars with $\tau_\ast \gg \tau_1$ and $\tau_o$ (also see
Mayor \& Mermilliod 1991 who use a different power index).  For G and
K dwarfs, $\tau_1 \sim 10^8$ yr and $V_1\sin i=V_r(\tau_1) \sin i \sim
30$ km s$^{-1}$.  However, the F dwarfs in the Hyades lose very little
angular momentum through stellar winds and $V_r \sin i$ continues to
extend up to 100 $\kms$ at $\tau_\ast = 6 \times 10^8$ yr.  Thus
$V_o$, $V_1$, $\tau_o$, $\tau_1$, and the power indices in eqs
(\ref{eq:vrot}) and (\ref{eq:vrot1}) may be functions of $M_\ast$.

From the empirical equation (\ref{eq:vrot}), we deduce that
\begin{equation}
\dot \omega_\ast = {\dot V_r \over R_\ast} 
\simeq - \frac{\gamma}{2}\left(
{\Omega_o \over \tau_o} \right) \left( {\Omega_\ast \over \Omega_o} \right)^3
\simeq - {\gamma V_o \over 2 R_\ast
\tau_o} \left( {\tau_\ast \over  \tau_o} \right)^{-1.5} 
\label{eq:omegas}
\end{equation} 
where differentiation is with respect to $\tau_\ast$, $\Omega_o =
V_o/R_\ast$ and $\gamma$ is a calibration factor which is set to unity
for G and K dwarfs. Eq(\ref{eq:vrot1}) yields a similar expression:
\begin{equation}
\dot \omega_\ast 
\simeq - {\gamma V_o \over 2 R_\ast
\tau_o} \left( \left({ V_o \over V_1 \sin i} \right)^2 + 
{\tau_\ast -\tau_1 \over  \tau_o} \right)^{-1.5} .
\label{eq:omegas2}
\end{equation} 
In both expressions, the timescale for spin-down
induced by a stellar wind is
\begin{equation}
\tau_{\omega_\ast} \equiv V_r /
\vert \dot V_r \vert \simeq (2 \tau_o / \gamma) (\Omega_o/\Omega_\ast)^2.
\end{equation}

Due to the initial rapid rate of angular momentum loss,
$\tau_{\omega_\ast} \ll\tau_{ \Omega_\ast}$ for rapidly rotating young
G and K dwarfs with $\tau_\ast $ greater than a few times $10^8$ yr
(i.e the wind-loss timescale is much shorter then the tidally driven
synchronization timescales). The evolution of the spin rates of these
stars is mainly determined by eq(\ref{eq:vrot}) rather than
eq(\ref{eq:omedot}). However, after these have declined to
sufficiently small values, the establishment of a quasi-equilibrium
state is possible in which the rate of angular momentum loss via a
stellar wind (eq \ref{eq:jdo}) is balanced by the rate at which the
star gains angular momentum from the orbit as the planet attempts to
spin up the star. Under such circumstances, $\dot\Omega_\ast=0$ so
that from eqn~\rn{eq:omedot} an equilibrium spin rate \be
\Omega_e=\frac{f_3(e)}{f_4(e)}n_e+
\frac{2}{9}\left(\frac{\alpha_\ast\epsilon_\ast
Q_\ast^\prime}{f_4(e)n_e}\right)
\left(\frac{M_\ast}{M_p}\right)^2
\left(\frac{a_e}{R_\ast}\right)^3\dot\omega_\ast
\label{eq:oequi}
\ee is possible provided that \be
|\dot{J}_{\omega_\ast}|<\frac{2}{7}\frac{f_3(e)}{\sqrt{1-e^2}}\,\frac{J_o}{\tau_{ep}}.
\ee In eqn~\rn{eq:oequi}, $a_e$ is the equilibrium semi-major axis and
$n_e=n(a_e)$ is the associated mean motion. This quasi-equilibrium can
be established through the evolution of either $\Omega_\ast$ or $a$ so
that it is attainable on a timescale given by the minimum of
$\tau_{\omega_\ast}$, $\tau_{\Omega_\ast}$, and $\tau_a$.  For a given
total angular momentum, if $\Omega_p =n$ then the equilibrium
quantities of $a_e$ and $\Omega_e$ can be determined self-consistently
from eqs~(\ref{eq:oequi}) and (\ref{eq:jtotal}) as functions of $e$.
\be J_{\rm total} (a_{e}, \dot \omega_\ast, e)= a_{e}^{9/2} A(e,\dot
\omega_\ast) + a_{e}^{1/2} B(e) + a_{e}^{-3/2} C(e)
\label{eq:aequi}
\ee where 
\be A(e,\dot \omega_\ast) =
\frac{2\alpha_{\ast}^{2}\epsilon_{\ast}Q_{\ast}^{\prime}
M_{\ast}^{3}\dot \omega_\ast}{9 f_{4}(e)
\sqrt{GM_{\ast}}M_{p}^{2}R_{\ast} }, \ee \be B(e) = M_{p} \left(
GM_{\ast}\left( 1 - e^{2} \right) \right)^{1/2}, \ee and \be C(e) =
\alpha_{p} M_{p} R_{p}^{2} \sqrt{GM_{\ast}} + \alpha_{\ast}
\epsilon_{\ast} M_{\ast} R_{\ast}^{2} f_{3}(e) \sqrt{GM_{\ast}} /
f_{4}(e) .  \ee

In Figure~\ref{figure4}(a), we plot $a_e$ and $\Omega_e$ as functions
of $J_{total}$ for three values (0, 0.2, 0.5) of the
eccentricity. The total angular momentum is normalized by $J_{0.1
AU}$, the value of Jupiter's angular momentum in a $0.1 AU$ orbit
around the sun. Similar to the calculation presented in Figure 2, we
adopt $M_\ast = 1 M_\odot$, $R_\ast = 1 R_\odot$, $M_p = 1 M_J$, $R_p
= 1.35 R_J$, $Q_p^\prime = Q_\ast ^\prime = 10^6$, $\alpha_\ast
\epsilon_\ast = 0.01$, and $\alpha_p \epsilon_p =0.2$. However, the
establishment of this equilibrium for a wide range in $a$ is only
possible for a very slow rate of spin-down, so we set $\dot
\omega_\ast$ to be $10^{-4}$ that in eq(\ref{eq:omegas}). In
Figure~\ref{figure4}(b), we represent a slower spin down rate with the
same model parameters except we set $\omega_\ast$ to be one tenth that
in Figure~\ref{figure4}(a).

The results in Figures~\ref{figure4}(a)and~\ref{figure4}(b) show that
$a_e$ is sensitive to both $e$ and $J_{total}$, while $\Omega_e$
varies primarily with $J_{total}$. The structures of the curves are
indicative of the dominant source of angular momentum in the system,
and thus the behavior of the equilibrium is highly sensitive to the
planets position. For low eccentricity systems with $a_e > 0.1$, the
dominant source of angular momentum is the orbit. Because $\dot
J_{total} = \dot J_{\dot \omega} < 0$, angular momentum is
continually drained from the system, but this only requires minute
adjustments to $a_e$.  For low eccentricity systems with $a_e < 0.1$AU
the stellar spin is more prominent and in order to maintain this spin,
larger changes in $a_e$ are necessary.  When $a_e<0.1$AU, a planet will
move closer to the star as its orbit is circularized, and this will be
accompanied by the spin-up of the star.  Based on these results (with
the planetary and stellar masses and radii given earlier) we infer
that as angular momentum is continually lost via stellar winds, the
state of near synchronization (in which $\Omega_p = n \simeq
\Omega_\ast$ and $e \simeq0$) occurs at a semi-major axis of
approximately 0.02-0.06 AU and $J_{total} / J_{0.1 {\rm AU}} \approx 1.0$,
corresponding to a total angular momentum of approximately
$10^{49}$g cm$^{2}/s$. It is approached asymptotically as $J_{total}$
slowly decreases.

\subsection{Present Spins of Host Stars}
In general there are two classes of quasi-angular momentum equilibria.
The close equilibria with $a_e < a_c$ are unstable to tidally induced
angular momentum exchange because they enhance the difference between
$\Omega_\ast$ and $n$ and lead to further orbital evolution (Hut 1980,
Lin 1981, Rasio {\it et al.} 1996).  There are also those distant
equilibria with $a_e > a_c$ which are stable against tidal interaction
because they lead to the reduction of difference between $\Omega_\ast$
and $n$.

The magnitude of $a_c$ does not explicitly depend on $J_{total}$ (see
eq \ref{eq:acrit}).  However, when the stellar spin equilibrium is
attained, its magnitude $a_{c0}$ is determined by $\Omega_\ast$ and
$n$, which along with $a_e$ and $J_{total}$ are functions of $e$.  For
the above model parameters, $a_{c0} = 0.029$AU for $e=0.2$.  We plot
the value of $a_{c0}$ and the corresponding $J_{total}$ for various
values of $e$ in Figures~\ref{figure4}(a)and~\ref{figure4}(b).  

The results in these figures also show that $a_e$ is a decreasing
function of $J_{total}$ (eq \ref{eq:aequi}).  In the large $J_{total}$
limit, if most of the angular momentum is retained by the planet's
orbit, $a$ would be large and the rate of tidal transfer of angular
momentum would generally be too small to balance the spin down induced
by the stellar wind. However, if a large amount of angular momentum is
stored in the spin of the planets and their host stars, a small
$a_e$ would be sufficient to
balance against the large magnitude of $\dot J_\omega$.  For modest
values of $J_{total}$, it is possible for angular momentum to be mostly
contained in the planet's orbit and for the host star to spin
relatively slowly with a small magnitude of $\dot J_\omega$.  In this
case, a spin equilibrium can be attained with modest values of $a_e$.
Note that there is an upper limit to the magnitude of $a_e$ ($\sim
0.2-0.3$AU), larger than which the rate of tidal transfer of angular 
momentum cannot be balanced with the wind loss rate.  

The evolutionary fate of planets depends on the magnitude of
$J_{total}$ when $a$ reaches $a_e$.  Consider the possibility that a
planet has migrated, through its interaction with a nascent disk of
gas, residual planetesimals, or other planets, close to the surface of
a rapidly spinning host star. If the planet is deposited in a region where $\Omega_\ast$ is smaller then $n$ the planet will not survive. Transfer of angular momentum from the orbit to the stellar spin will cause the planet to migrate into the star. However, if the planet is deposited outside co-rotation ($\Omega_\ast > n$) its orbit will expand through star-planet tidal interaction while the star spins down through wind losses until the system comes into equilibrium at $a=a_e$. As the star continues to loose angular momentum through its winds, the total angular momentum of the system will decrease and the location of $a_e$ will move inward, as can be seen in Figures ~\ref{figure4}(a) and ~\ref{figure4}(b). If the system still retains a relatively large $J_{total}$ (stored primarily in the stellar spin) the magnitude angular momentum lost through winds will be too great for the system to remain in equilibrium. The planet's inward migration would continue until it merges with its host star on a time scale $\tau_a (a_e) < \tau_{\Omega \ast} (a_e)$ which, for planets with $P < 2$ days, is less than a few Gyr. However, if a spin equilibrium is retained after $J_{total}$ has already reduced to modest values such that $a_e > a_{c0}$, it would be stablized against further evolution.  Note that the values of $a_{c0}$ correspond to orbital periods in the range of 3-4 days, which are close to the observed cut off in the period distribution of extra solar planets.

In the previous subsection, we indicated that main sequence stars
continually lose angular momentum through stellar winds. For example,
the ultra fast rotators among the young G and K dwarfs spin down
rapidly.  During their spin down, the $\Omega_\ast$ of these stars may
become comparable to the $n$ of their short-period planets.  But this
state of near synchronization cannot be maintained because the rate of
tidal transfer of angular momentum between a star and its planet,
$\dot J_\ast$, is inadequate to balance that of wind-driven loss of
the stellar spin angular momentum $\dot J_{\dot
\omega}$. Consequently, $\Omega_\ast$ reduces to $\Omega_e < < n$ and
$a$ declines through a sequence of non-synchronized (with $\Omega_\ast
< \Omega_s \simeq n $) quasi-equilibria on the timescale of $\tau_a$.

This evolutionary path may have been taken by some G and K dwarf host
stars of the shortest-known-period ($\sim 3$ days) planets such as HD
46375 (Marcy {\it et al.} 2001), HD~177949 (Tinney {\it et al.} 2001),
and HD~187123 (Butler {\it et al.} 1999).  The corresponding $V_r$ for
synchronous rotation (with $\Omega_\ast = n$) in these systems is
$\sim 15$ km s$^{-1}$ which is comparable to that of typical dwarf
stars in the $\alpha$ Persei and Pleaides clusters (assuming the
expectation value $< \sin i > = 0.7$).  It is entirely possible that
some time in the past, the spins of these stars may have been
synchronous with the mean motion of their planets.  With their present
configurations and $Q_\ast^\prime = 10^6$, their present $\tau_{\Omega
\ast} $ of a few Gyr is comparable to their estimated $\tau_\ast$.
Yet the spin of these stars is slow ($\Omega_\ast < n$), as indicated
by their quite chromosphere.  We suggest that the effect of the tidal
interaction is overshadowed by that of the stellar wind in these G and
K dwarfs when they evolved through a state of synchronization.
Although they have attained a stable quasi-equilibrium today with $a
\sim 2 a_c$, their orbit may decay into an unstable configuration with
$a < a_c$.  In order for these planets to have survived $\tau_\ast
\sim$ a few Gyr, their $\tau_a \sim 2 \tau_{\Omega \ast} > \tau_\ast$
which requires $Q_\ast^\prime$ to be comparable to or greater than
$10^6$.  In principle, a nearly synchronized quasi-equilibrium (with
$\Omega_\ast \simeq \Omega_s \simeq n$) is attainable for the
longer-period (with $P > 10$ days) planets around G and K dwarfs, but
the timescale for evolution into such a state is much longer than the
age of the mature stars ({\it i.e.} $\tau_{\Omega \ast} > >
\tau_\ast$) which makes their potential existence coincidental.

But for the more massive F dwarfs, $\tau_{\Omega \ast} < \tau_{\omega
\ast}$ because their $\dot \omega_\ast$ is much smaller than that of
the G and K dwarfs.  For these stars, we can use eq (\ref{eq:omegas})
for $\dot \omega_\ast$ by setting $\gamma \sim 0.1$.  In this case,
the state of quasi-equilibrium may be maintained because the
transitory decline of $\Omega_\ast$ can be compensated by a tidal
transfer of angular momentum from the planet's orbit to the stellar
spin, resulting in an net increase in both $\Omega_e$ and $n (a_e)$
and a reduction in $a_e$. In addition, the relatively shallow
convection zone of the F dwarfs also reduces their $\epsilon_\ast$ and
$a_c$ well below that of the G and K dwarfs.  It is thus possible to
attain a sequence of nearly synchronized (with $\Omega_\ast \simeq
\Omega_s \simeq n$) quasi equilibria with $a_e > a_c$.  This sequence
of events may have led to the present orbital configuration of the
planet around $\tau$ Boo.

\subsection{Numerical Models of Spin-Orbit Evolution}
In order to verify the above qualitative and analytic description, we
numerically integrate, with a fourth-order Runge-Kutter scheme, eqs
(\ref{eq:edot}), (\ref{eq:omedot}), (\ref{eq:omepdot}), and
(\ref{eq:adot}) with appropriate substitutions for $f_{1, 2, 3, 4}$,
$\dot \omega_\ast$, $\dot J_{p, \ast}$.  The dependent variables are
$\Omega_{p, \ast}, a,$ and $e$.  We adopt the standard model
parameters in which $M_\ast = 1 M_\odot$, $M_p = 1 M_J$, $R_\ast = 1
R_\odot$, $R_p = 1.5 R_J$, $\dot \omega_p =0$, and $Q_p ^\prime
=10^6$. Assuming that $Q_\ast^\prime$ is around $1.5\times10^5$ before
$t=10^8$ yr and $10^6$ after $t=10^9$ yr, the star's dissipation
parameter is taken to be of the form $Q_\ast^\prime= a+b\, {\rm
erf}[(\tau^\prime-5)/3]$, where $\tau^\prime = t/ 10^8$~yr, $a=5.75
\times 10^5$ and $b=4.2\times 10^5$.

In order to demonstrate several possible outcomes, we constructed six
models.  In models 1 and 2, we set $P$ = 3 days at $t=0$ to represent
the shortest period planets.  In models 3 and 4, $P$ is set to be 7
days at $t=0$ as a possible initial condition of borderline planets.
In models 5 and 6, we choose $P = 1.5$ days to represent some
hypothetical ultra-short-period planets which may have once existed
close to the surface of their host stars. To explore the effects of
stellar spin, models 1, 3, and 5 start with $\Omega_\ast = 10 n$,
while in models 2, 4, and 6, $\Omega_\ast = 9 n$. These spin rates are
rapid, but not near the break up values. In all models, we choose
$\Omega_p = 10 n$ and $e =0.2$ at $t=0$. All models are integrated for
$6$ Gyr.

In Figure~\ref{figure5}(a), we plot the $e$ evolution of models 1, 2,
3, and 4 which are represented with solid, short-dashed, long-dashed,
and dot-dashed lines respectively.  Both models 5 and 6 are
represented by dotted lines.  In
Figures~\ref{figure5}(b),~\ref{figure5}(c), and~\ref{figure5}(d), we
plot the evolution of $a$, $\Omega_p/n$ and $\Omega_\ast/n$
respectively.

After brief initial episodes of $e$ excitation, $e$ declines rapidly
in models 1, 2, 5, and 6. Close to the star, $\tau_e \approx 3$ Gyr
for the short-period planets, and rapid late stage $e$ damping erases
any trace of early $e$ excitation. Despite an initial decline in the
eccentricities in models 5 and 6, the planet migrates outward and $e$
levels off. Eventually it is caught at in the quasi-equilibrium at
$a_e$, and $e$ begins to drop again. In models 3 and 4, the
eccentricity is maintained at non-zero values even after $6$ Gyr. The
contrasts seen in the models with different stellar spin rates
supports our conjecture that the dispersion in the $e-P$ distribution
may be due to tidal dissipation in host stars with various rotation
speeds.

The planets quickly decrease in spin frequency due to the artificially
high initial values. They are all able to attain synchronous spin
frequency within $\sim1-2$ Gyr, with the closer planets doing so more
quickly.  The effect of the stellar spin on planetary spin is
negligible.

After $400-500$ Myr, all the models are caught in the
quasi-equilibrium described earlier in this paper. As angular momentum
is lost to stellar-winds it is balanced by tidal transfer from the
orbit to the star. Both the semi-major axis and stellar spin reach
their equilibrium values described by eq (\ref{eq:oequi}), eq
(\ref{eq:aequi}) and shown in Figure~\ref{figure4}. Models 2, 4, and
6, that start with slower initial stellar spins, have correspondingly
smaller equilibrium stellar spins and semi-major axes. Any subsequent
evolution proceeds on much longer timescales governed by $\dot
\omega_{\ast}$.

In all the models the stellar spin steadily decreases from a large
initial value. However, Figure \ref{figure5}(d) reflects changes in
both stellar spin and semi-major axis. The stellar spin in model 1
initially decreases more quickly then the orbital frequency, but by
$\sim 100$ Myr the orbital frequency is decreasing faster and
$\Omega_{\ast}/n$ increases. The same behavior occurs in model 5, but
at an earlier time. Models 3 and 4 have such large semi-major axes
that the stellar spin must decrease significantly before any
equilibria can be established. In Models 3, 4, 5, and 6 the stellar
spin has stopped changing by the time the semi-major axis reaches its
equilibrium value. In contrast, the stellar spin in models 1 and 2
continues to decrease for about $100$ Myr, even after the semi-major
axis has stopped changing significantly. Further evolution should
slowly move the system from quasi-equilibrium into synchronization
($\Omega_p = n \simeq \Omega_\ast$ and $e \simeq0$).

\section{Summary and Discussion}

In this paper we have considered the effect of tidal interaction
between main sequence dwarf stars and their short-period planets.  It
is generally customary to assume that the orbital eccentricity of
planets is primarily damped by the dissipation of stellar tidal
disturbance inside the planets.  Such a scenario provides a reasonable
account for the circular orbits of planets with periods less than
about a week.  But for planets with periods slightly longer than a
week, there is a spread of orbital eccentricity.  We suggest that this
dispersion is associated with different past rotational speed and a
range of spin down rate among their host stars.  This conjecture can
be best verified by the detection of short-period planets around stars in young
clusters such as $\alpha$ Persei and Pleiades.  We expect short-period
planets around rapidly rotating F, G, and K dwarfs to have relatively
large eccentricities.  For older clusters such as the Hyades,
short-period planets around F dwarfs should have larger eccentricities
than those around G and K dwarfs, mainly due to the difference in the
stellar spin-down rate.

Rapid stellar spin also causes outward orbital migration on slightly
longer timescales.  Thus we expect the minimum orbital period for planets around
rapidly rotating dwarfs in young clusters to be less than its present 3 day
value.  In contrast, this cutoff period for slowly rotating dwarfs in
young clusters should be the same as its present value.  It is also
possible that the period cutoff represents the boundary corresponding
to survivability against tidally induced orbital decay.  In that case,
the period cutoff for slowly rotating dwarfs in young clusters would
be shorter than 3 days.

During their spin down, the stars' spin angular frequency temporarily
coincides with the mean motion of the planets.  Among the short-period
systems, this state of near synchronous rotation can only be preserved
around F dwarfs which spin down gradually.  All main sequence dwarfs
attain a quasi-angular momentum equilibrium in which the host stars'
loss of angular momentum through stellar winds is balanced by their
tidal transfer of angular momentum with their planets.  For G and K
dwarfs, the equilibrium frequencies are much smaller than that for F
dwarfs.  Consequently, it is much easier to form short-period nearly
synchronous systems around F dwarf than around G and K dwarfs.

Despite the enormous progress that has been made in the area of
extra solar planet searches, there is insufficient data to verify out
conjecture.  But future radial-velocity, astrometric, and transit
searches for planets in clusters of various ages should provide
sufficient data to test the scenario presented here.

We thank D. Fischer, P.G. Gu, S. Martell, and L. Langland-Shula for
useful conversations.  D. Lin thanks the director, D.O. Gough, of the
Institute of Astronomy, Cambridge University for hospitality while
some of this work was completed.  This research has been supported in
part by the Sackler Foundation, NSF through grants AST-9987417, by
NASA through grants NAG5-7515, NAG5-8196, and NAG5-10727, an
astrophysics theory program which supports a joint Center for Star
Formation Studies at NASA-Ames Research Center, UC Berkeley, and UC
Santa Cruz, and by the Victorian Partnership for Advanced Computing.

\clearpage

\begin{reference}

\reference Artymowicz, P. 1992, \pasp, 104, 769

\reference Bodenheimer, P., Lin, D.N.C.\ \& Mardling, R.A. 2001,
\apj, 548, 466

\reference Brown, T. M., Charbonneau, D., Gilliland, R. L., Noyes, 
R. W., Burrows, A. 2001, \apj522, 699

\reference Burrows, A., Guillot, T., Hubbard, W. B., Marley, M. S., 
Saumon, D., Lunine, J. I. \& Sudarsky, D. 2000, \apj, 534, 97

\reference Burrows, A., Marley, M., Hubbard, W. B., Lunine, J. I., 
Guillot, T., Saumon, D., Freedman, R., Sudarsky, D.\ \& Sharp, C. 
1997, \apj, 491, 856

\reference Butler, R. P., Marcy, G. W., Vogt, S. S., Apps, K.
1998, PASP, 110, 1389

\reference Butler, R.P., Marcy, G.W., Fischer, D.A., Brown, T.M.,
Contos, A.R., Korzennik, S.G., Nisenson, P. \& Noyes, R.W.
1999, \apj, 526, 916

\reference Cowling, T. G. 1941, \mnras, 101, 367

\reference Dintrans, B., \& Ouyed, R. 2001, \aap, 375, L47

\reference Eggleton, P.P., Kiseleva, L.G., Hut, P.\ 1998,
\apj, 499, 853

\reference
Fischer, D.A., Marcy, G.W., Butler, R.P., Vogt, S.S.\ \& Apps, K.\ 1999
PASP, 111, 50

\reference Goldreich, P., \& Nicholson, P. D. 1977, Icarus, 30, 301

\reference Goldreich, P., \& Nicholson, P. D. 1989, \apj, 342, 1079

\reference Goldreich, P.\ \& Sari, R.\ 2003, \apj, 585, 1024

\reference Goldreich, P.\ \& Soter, S.\ 1966, Icarus, 5, 375

\reference Goldreich, P., \& Tremaine, S. 1980, \apj, 241, 425

\reference Goodman, J.\, \& Oh, S.\ P.\ 1997, \apj, 486, 403

\reference Gu, P.G., Lin, D.N.C., \& Bodenheimer, P.  \
2003, \apj, 588, 509

\reference Gu, P.G., Bodenheimer, P. \& Lin, D.N.C.  \
2004, \apj, 608, 1079

\reference Guillot, T., Hubbard, W. B., Stevenson, D. J., \&
Saumon, D. 2004, in Jupiter, ed. F. Bagenal, F., T. Dowling, \& W. McKinnonn (Cambridge: Cambridge Univ. Press), in press

\reference Hut, P.\ 1980, Astron. Astrophys., 92, 167

\reference Iben, I. 1965, \apj, 141, 993

\reference Iben, I. 1967, \apj, 147, 624

\reference Ioannou, P. J., \& Lindzen, R. S. 1993a, \apj, 406, 252

\reference Ioannou, P. J., \& Lindzen, R. S. 1993b, \apj, 406, 266

\reference Ioannou, P. J., \& Lindzen, R. S. 1994, \apj, 424, 1005

\reference Lin, D. N. C.\ 1981, MNRAS, 197, 1081

\reference Lin, D. N. C., Bodenheimer, P., \& Richardson,
D. C. 1996, \nat, 380, 606

\reference Lubow, S. H., Tout, C. A., \& Livio, M. 1997, \apj, 484, 866

\reference Marcy, G. W., Butler, R. P., \& Vogt, S. 2000a, \apj, 536, 43

\reference Marcy, G. W., Butler, R. P., Williams, E.,
Bildsten, L., Graham, J. R., Ghez, A. M., \& Jernigan, J. G. 1997, 
\apj, 481, 926

\reference Marcy, G.~W., Cochran, W~.D., \& Mayor, M. 2000,
in Protostars and
Planets IV, ed. V. Mannings, A.~P. Boss \& S.~S. Russell
(Tucson:Univ. of Arizona Press), 1285

\reference Mardling, R.A.\ \& Lin, D.N.C.  2002, \apj, 573, 829

\reference Mathieu, R. D. 1994, \araa, 32, 465

\reference Mayor, M. \& Mermilliod, J.C. 1991 in Angular 
Momentum Evolution of Young Stars, eds S. Catalano \& J.R. Stauffer,
(Dordrecht: Kluwer), 143

\reference Murray, C.\ D.\ \& Dermott, S.\ F.\ 1999, Solar System Dynamics,
Cambridge University Press, Cambridge

\reference Murray, N., Hansen, B., Holman, M. \& Tremaine, S. 1998,
Science, 279, 69

\reference Ogilvie, G.I.\ \& Lin, D.N.C. 2004, \apj, 610, 477

\reference Papaloizou, J.C.B., Nelson, R.P. \& Masset, F. 2001, Astron. 
Astrophys., 366, 263

\reference Papaloizou, J. C. B., \& Savonije, G. J. 1997, \mnras, 291, 651

\reference Peale, S.\ J.\ 1999, \araa, 37, 533

\reference
Pepe, F., Mayor, M., Galland, F., Naef, D., Queloz, D., Santos, N., Udry, S.\ \& Burnet, M.\
2002, A\&A, 388, 632

\reference Prosser, C.F. 1992, \aj, 103, 488

\reference Rasio, F. A., \& Ford, E. B. 1996, Science, 274, 954

\reference Rasio, F. A., Tout, C. A., Lubow, S. H., \&
Livio, M. 1996, \apj, 470, 1187

\reference Savonije, G. J., \& Papaloizou, J. C. B. 1983,
\mnras, 203, 581

\reference Savonije, G. J., \& Papaloizou, J. C. B. 1984, 
\mnras, 207, 685

\reference Savonije, G. J., Papaloizou, J. C. B., \& Alberts,
  F. 1995, \mnras, 277, 471

\reference Skumanich, A. 1972, \apj, 171, 565

\reference Soderblom, D., Stauffer, J., Hudon, J.D.\ \& Jones, B. F.\ 1993a \apjs, 85, 315

\reference Soderblom, D., Stauffer, J., MacGregor, K. B.\ \& Jones, B. F.\ 
1993b \apj, 409, 624

\reference Stassun, K., Mathieu, R., Mazeh, T.\ \& Vrba, F.\ 1999, \aj, 117, 2941

\reference Terquem, C.\ , Papaloizou, J.\ C.\ B.\ , Nelson, R.\ P.\ ,
\& Lin, D.\ N.\ C.\ 1998, \apj, 502, 788

\reference
Tinney, C. G., Butler, R. P., Marcy, G. W., Jones, H. R. A., 
Penny, A. J., Vogt, S. S., Apps, K., Henry, G.W. 2001, \apj, 551, 507

\reference Trilling, D.\ E.\ 2000, \apj, 537, 61

\reference Trilling, D.\ E.\ , Benz, W.\ , Guillot, T.\, Lunine, J.\ I.\ ,
Hubbard, W.\ B.\ , \& Burrows, A. 1998, \apj, 500, 428

\reference Vogt, S.S., Butler, R.P, Marcy, G.W., Fischer, D.A., Pourbaix, 
D., Apps, K.\ \& Laughlin, G.\ 2002, \apj, 568, 352

\reference Yoder, C. F., \& Peale, S. J. 1981, Icarus, 47, 1

\reference Zahn, J.-P. 1970, \aap, 4, 452

\reference Zahn, J.-P.\ 1977, \aap, 57, 383 

\reference Zahn, J.-P.\ 1989, \aap, 220, 112

\end{reference}

\begin{figure}[ht] 
\begin{center}
\includegraphics[scale=.90,angle=0,trim=0 0 0 0,clip=true]{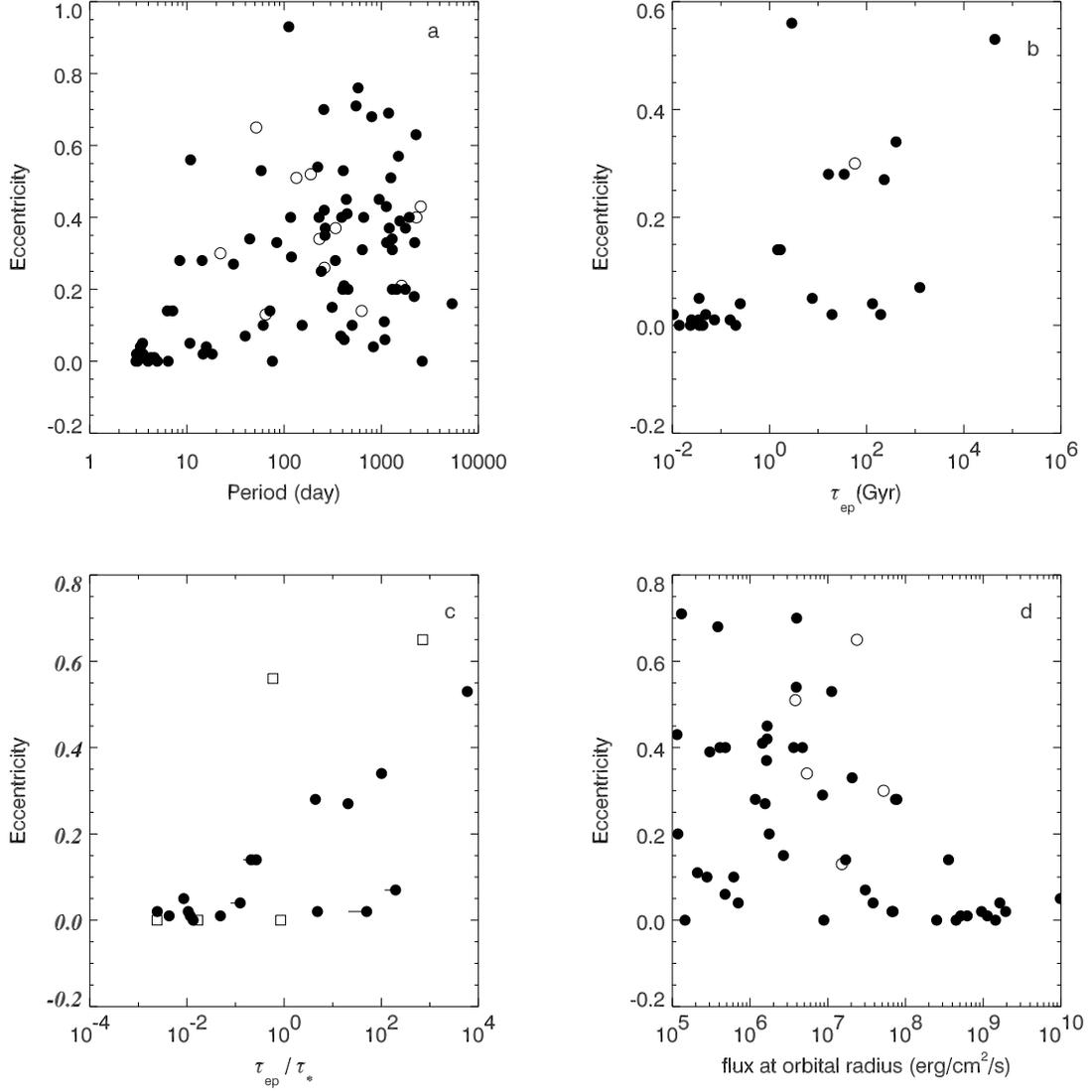}
\end{center}
\caption{\protect\small
(a) Period-eccentricity relation for extrasolar planets,
(b) Eccentricity as a function of eccentricity-damping timescale for planets
with orbital periods less than two months,
(c) Eccentricity as a function of eccentricity-damping timescale normalized by the stellar age,
(d) Eccentricity as a function of stellar flux received at planets orbital radius.
}
\label{figure1}
\end{figure}

\newpage
\begin{figure}[ht] 
\begin{center}
\includegraphics[scale=.80,angle=0,trim=0 0 0 0,clip=true]{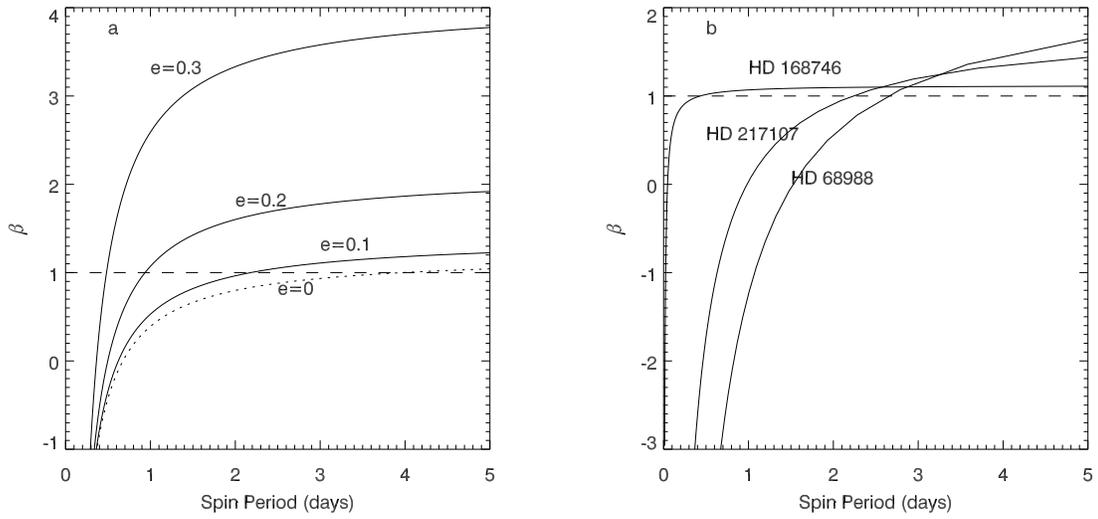}
\end{center}
\caption{\protect\small
Eccentricity damping efficiency factor $\beta$ vs.\ stellar spin period.
(a) Sun/Jupiter parameters with $Q_\ast=Q_p$, $P=6.5$ days and
various eccentricities.
(b) Three representative borderline systems. 
}
\label{beta}
\end{figure}
\newpage
\begin{figure}[ht] 
\begin{center}
\includegraphics[scale=.90,angle=0,trim=0 0 0 0,clip=true]{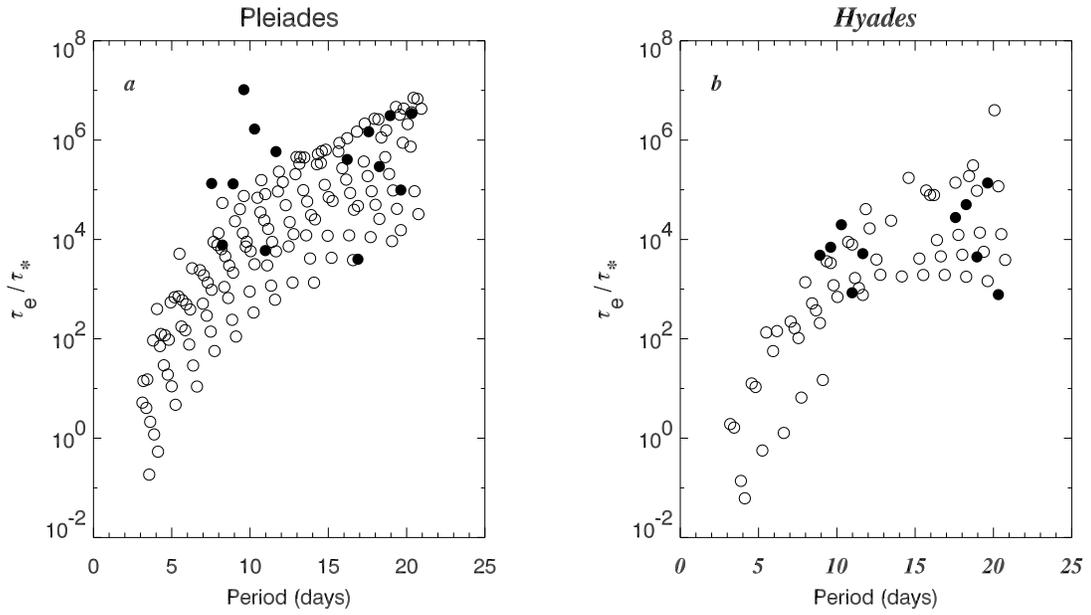}
\end{center}
\caption{\protect\small
$\tau_{e}/\tau_{\star}$ for stars as a function of orbital period. Empty circles correspond to eccentricity damping and filled circles are excitation timescales. (a) Pleiades cluster ($\tau_{\star} = 70 Myr$).
(b) Hyades cluster ($\tau_{\star} = 600 Myr$). $\tau_{e}$ was calculated assuming planet-orbit synchronization and eccentricities between 0.0 and 0.7 (see text for more details).
}
\label{figure3}
\end{figure}
\newpage
\begin{figure}[ht] 
\begin{center}
\includegraphics[scale=.90,angle=0,trim=0 0 0 0,clip=true]{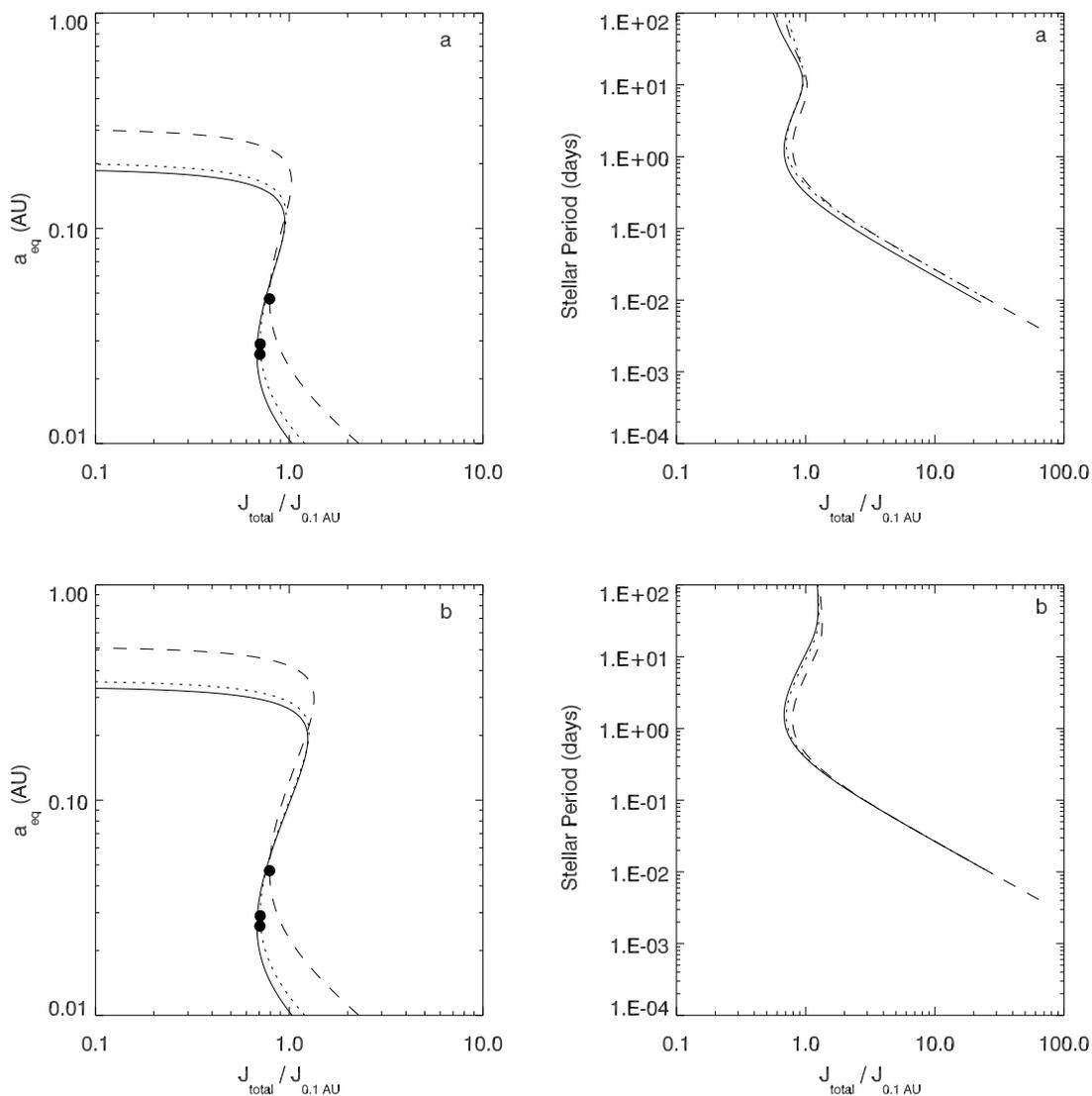}
\end{center}
\caption{\protect\small
(a) The equilibrium semi-major axis and stellar spin period as a function of total angular momentum in the system for three eccentricities: e=0 (solid), e=0.2 (dotted), and e=0.5 (dashed). The equilibrium is reached when the angular momentum lost to stellar winds is balanced by that transfered from the orbit and spin of the planet such that the stellar spin remains constant. (b) Stellar wind set at one-tenth that in (a). 
}
\label{figure4}
\end{figure}
\newpage
\begin{figure}[ht] 
\begin{center}
\includegraphics[scale=.90,angle=0,trim=0 0 0 0,clip=true]{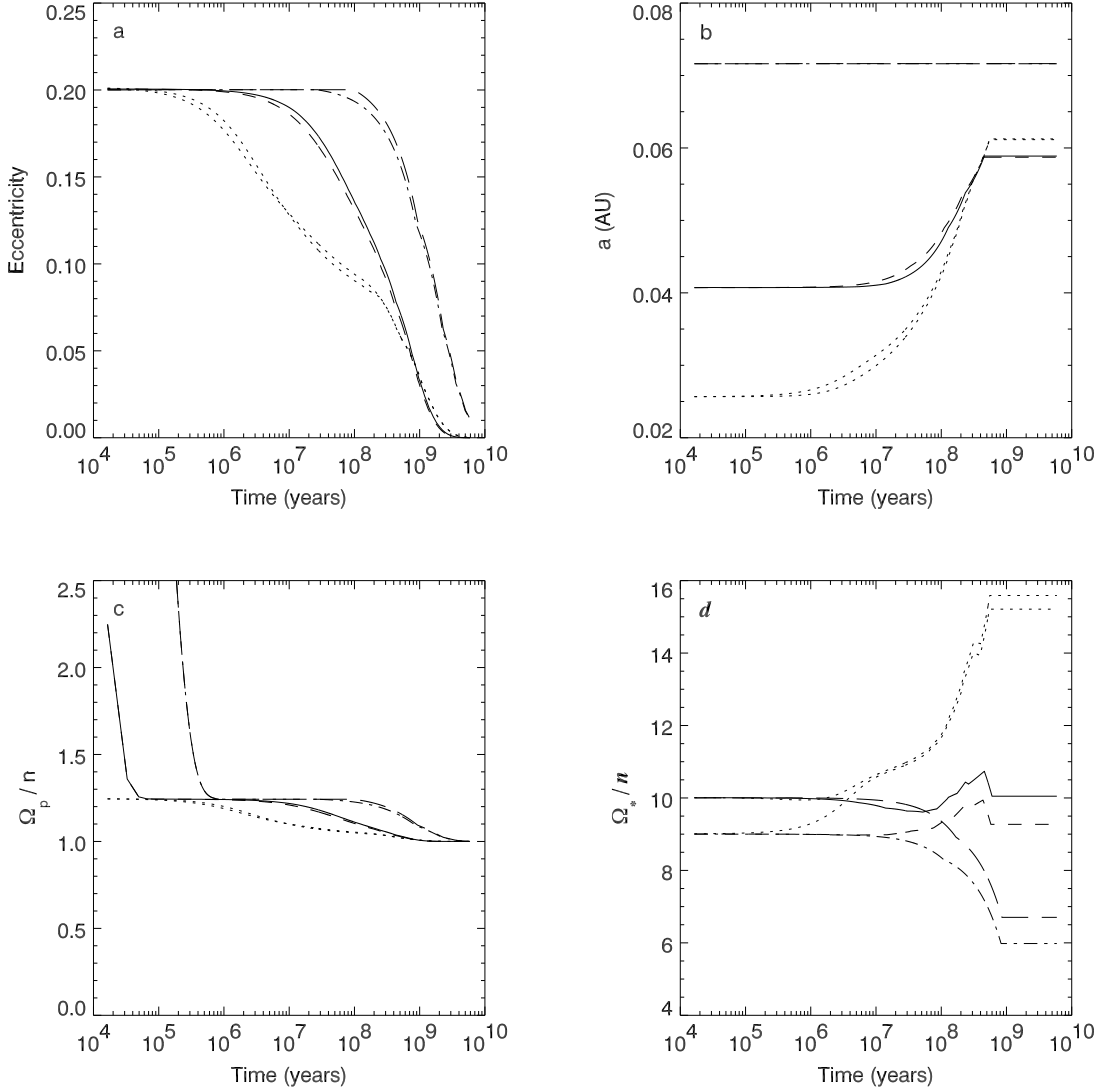}
\end{center}
\caption{\protect\small
Results of a long-term integration of eccentricity, semi-major axis, stellar spin frequency and planetary spin frequency. Models 1, 2, 3 , and 4 are represented with solid, short-dashed, long-dashed, and dot-dashed lines respectively. Models 5 and 6 are represented with dotted lines.
(a) Eccentricity as a function of time,
(b) Semi-major axis as a function of time,
(c) Ratio of planetary spin to orbital frequency as a function of time,
(d) Ratio of stellar spin to orbital frequency as a function of time.
}
\label{figure5}
\end{figure}

\end{document}